\documentclass[showpacs,10pt,twocolumn,prb]{revtex4-1}

\usepackage{amsmath}
\usepackage{amssymb}
\usepackage{graphicx}
\usepackage{amssymb}
\usepackage{graphics}
\usepackage{epsfig}
\usepackage{CJK}
\usepackage{color}

\begin{document}

\title{Sustained phase separation and spin glass in Co-doped K$_{x}$Fe$_{2-y}$Se$_{2}$ single crystals}
\author{Hyejin Ryu,$^{1,2,\dag}$ Kefeng Wang,$^{1,\S}$ M. Opacic,$^{3}$ N. Lazarevic,$^{3}$ J. B. Warren,$^{4}$ Z. V. Popovic,$^{3}$ Emil S. Bozin$^{1}$ and C. Petrovic$^{1,2}$}
\affiliation{$^{1}$Condensed Matter Physics and Materials Science Department, Brookhaven National Laboratory, Upton, New York 11973, USA\\
$^{2}$Department of Physics and Astronomy, Stony Brook University, Stony Brook, New York 11794-3800, USA\\
$^{3}$Center for Solid State Physics and New Materials, Institute of Physics Belgrade, University of Belgrade,
Pregrevica 118, 11080 Belgrade, Serbia\\
$^{4}$Instrument Division, Brookhaven National Laboratory, Upton, New York 11973, USA}

\date{\today}

\begin{abstract}

We present Co substitution effects in K$_{x}$Fe$_{2-y-z}$Co$_{z}$Se$_{2}$ (0.06 $\leq$ $z$ $\leq$ 1.73) single crystal alloys. By 3.5\% of Co doping superconductivity is suppressed whereas phase separation of semiconducting K$_{2}$Fe$_{4}$Se$_{5}$ and superconducting/metallic K$_{x}$Fe$_{2}$Se$_{2}$ is still present. We show that the arrangement and distribution of superconducting phase (stripe phase) is connected with the arrangement of K, Fe and Co atoms. Semiconducting spin glass is found in proximity to superconducting state, persisting for large Co concentrations. At high Co concentrations ferromagnetic metallic state emerges above the spin glass. This is coincident with changes of the unit cell, arrangement and connectivity of stripe conducting phase.

\end{abstract}

\maketitle

\section{Introduction}

Since the discovery of high temperature Fe-based superconductivity,\cite{KamiharaY} many types of Fe-based superconductors have been reported including K$_{x}$Fe$_{2-y}$Se$_{2}$.\cite{FujitsuS,GuoJ} Various novel phenomena were observed by chemical substitution on Fe site. For example, Co and Ni doping in FeAs tetrahedra of LaFeAsO and BaFe$_{2}$As$_{2}$-based pnictides give rise to superconductivity,\cite{SefatAS,MatsuishiS,LJA,SahaSR} whereas Co doping in FeSe suppresses superconductivity.\cite{MizuguchiY,KotegawaH} In particular, A$_{x}$Fe$_{2-y}$Se$_{2}$ (A = K, Cs, Rb, Tl) materials are strongly sensitive to chemical substitutions. \cite{YuY,LiMT,TanD}

Among several different types of Fe-based superconductors, A$_{x}$Fe$_{2-y}$Se$_{2}$ (A = K, Cs, Rb, Tl) materials generate significant attention due to unique characteristics such as the absence of the pocket in the Brillouin zone center and phase separation with Fe-vacancy order where crystal separates into (super) conducting stripes (block) and magnetic semiconducting matrix regions on (0.01-100)$\mu$m scale.\cite{QianT,WangZ,ZhangY,LiW,DingX,BaoW} The mechanism of the conducting and non-conducting states in proximity to K$_{x}$Fe$_{2-y}$Se$_{2}$ is of the great importance for the understanding of superconductivity.\cite{YinZP,DaiP} Consequently, the details of phase separation, phase stoichiometry, compositions as well as their magnetic and electric ground states are currently debated and are of high interest.

In this study we have investigated K$_{x}$Fe$_{2-y-z}$Co$_{z}$Se$_{2}$ (0.06 $\leq$ $z$ $\leq$ 1.73) single crystal alloys, where $y$ is Fe/Co vacancy. A rich phase diagram is discovered where phase separated superconducting state of K$_{x}$Fe$_{2-y}$Se$_{2}$ turns into spin glass and then to KCo$_{1.73}$Se$_{2}$ ferromagnetic metal with no phase separation. We show that microstructure arrangement and connectivity are rather important for ground state changes, in addition to changes induced by Co substitution for Fe.

\section{Experiment}

Single crystals of K$_{x}$Fe$_{2-y-z}$Co$_{z}$Se$_{2}$ (0.06 $\leq$ $z$ $\leq$ 1.73) were synthesized as described previously.\cite{LeiH107} Plate like crystals with size up to 10 $\times$ 10 $\times$ 3 $mm^{3}$ were obtained. High-energy medium resolution synchrotron x-ray experiment at 300 K was conducted on the X7B beamline of the National Synchrotron Light Source at Brookhaven National Laboratory. The setup utilized an x-ray beam 0.5 mm $\times$ 0.5 mm in size with a wavelength of 0.3196 Angstroms (E = 38.7936 keV) configured with a focusing double crystal bent Laue monochomator, and Perkin-Elmer amorphous silicon image plate detector mounted perpendicular to the primary beam path. Finely pulverized samples were packed in cylindrical polyimide capillaries 1 mm in diameter and placed 377.81 mm away from the detector. Multiple scans were performed to a total exposure time of 240 s. The two-dimensional (2D) diffraction data were integrated and converted to intensity versus 2$\theta$ using the software FIT2D.\cite{Hammersley} Structural refinements were carried out using the GSAS modeling program\cite{Larson} operated by the EXPGUI platform.\cite{Toby} The backscattered images and energy dispersive X-ray spectroscopy (EDX) mappings were performed in an JEOL-6500 scanning electron microscope (SEM). Electrical transport, thermal transport, heat capacity, and magnetization measurements were carried out in Quantum Design PPMS-9 and MPMS-XL5. Raman scattering measurements were performed on freshly cleaved (001)-oriented samples using TriVista 557 and Jobin Yvon T64000 Raman systems in backscattering micro-Raman configuration. The 514.5 nm laser line of a mixed Ar$^+/$Kr$^+$ gas laser was used as an excitation source. All measurements were carried out at room temperature, in the vacuum.

\section{Results and Discussion}

\begin{figure}
\centerline{\includegraphics[scale=0.2]{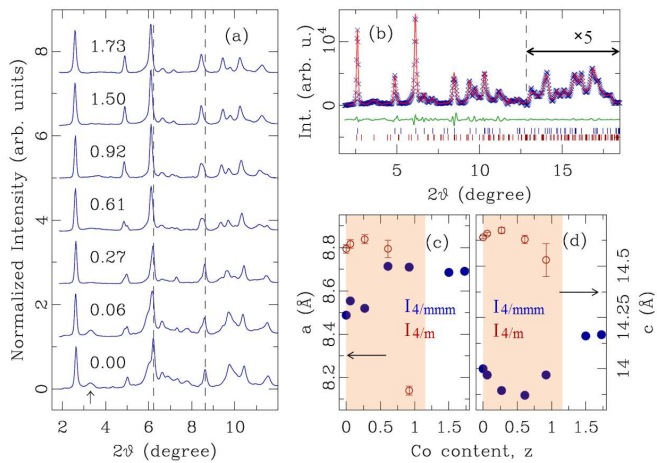}} \vspace*{-0.3cm}
\caption{(Color online)  (a) High-energy synchrotron x-ray diffraction data of the K$_{x}$Fe$_{2-y-z}$Co$_{z}$Se$_{2}$ series, normalized by the intensity of (002) reflection for comparison. Data are offset for clarity, and labeled by respective Co content as measured by EDX. Vertical dashed lines are provided as a visual reference. Vertical arrow indicates (110) reflection which is a hallmark of $I4/m$ phase. Coexistence of $I4/m$ and $I4/mmm$ phases can be visually tracked up to z=0.92(4) cobalt content in the samples studied. (b) Modeling of powder diffraction data for sample with z=0.92(4). Crosses represent data, red solid line is the model, green solid line is the difference which is offset for clarity. Vertical ticks mark the reflections in $I4/mmm$ (top row) and $I4/m$ (bottom row) phases. (c) and (d) display evolution with EDX-established Co content of refined lattice parameters for $I4/mmm$ (solid blue symbols) and $I4/m$ (open red symbols). All parameters are expressed in the $I4/m$ metrics (see text). Shaded is the region where the signatures of the phase coexistence could be reliably established from the diffraction data.}
\label{XRD}
\end{figure}

Obtained high-energy synchrotron XRD data of the K$_{x}$Fe$_{2-y-z}$Co$_{z}$Se$_{2}$ series [Fig. 1(a)] can be fitted very well with $I4/m$ and $I4/mmm$ space groups for z$\leq$0.92(4), while they are fitted by the $I4/mmm$ space group only for z$>$0.92(4). This implies coexistence of $I4/m$ and $I4/mmm$ phases when z$\leq$0.92(4). Typical fit for phase separated sample with z=0.92(4) is shown in Fig. 1(b). Notably for 0.27$\leq$z$\leq$0.92 intensities of reflections characteristic for $I4/m$ phase become appreciably weaker and rather broad - indicative of disorder and loss of structural coherence of this structural component, as well as its presumably diminishing contribution. However, quantitative phase analysis was not feasible due to the limited resolution of the measurement, and due to the diffuse nature of the signal with broad and overlapping reflections. Evolution of extracted lattice parameters with Co content is shown in Fig. 1(c) and (d). Lattice parameters for a-axis of the $I4/mmm$ space group are converted into comparable numbers for the $I4/m$ space group using the formula $I4/m = \sqrt{5}$ $I4/mmm$. Nonmonotonic evolution of lattice parameters highlights the complex crystal structure and bonding in K$_{x}$Fe$_{2-y-z}$Co$_{z}$Se$_{2}$.

\begin{figure}
\centerline{\includegraphics[scale=0.4]{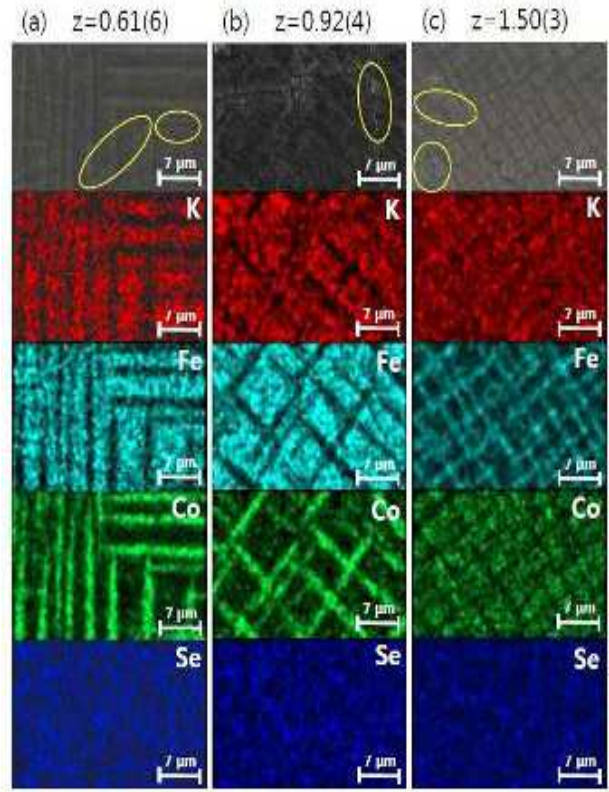}} \vspace*{-0.3cm}
\caption{(Color online) Back-scattered electron images of SEM measurement and EDX mappings of K$_{x}$Fe$_{2-y-z}$Co$_{z}$Se$_{2}$ when (a) $z=0.61(6)$, (b) $z=0.92(4)$, and (c) $z=1.50(3)$.}
\label{SEM}
\end{figure}

The surface morphologies (Fig. 2) show that the Co-doped crystals separate into two regions, stripe(domain)-like brighter area with 1$\sim$2 $\mu$m thick and darker matrix area, similar to pure K$_{x}$Fe$_{2-y}$Se$_{2}$,\cite{DingX} implying that the phase separation is preserved with Co doping. Distributions of the elements of K$_{x}$Fe$_{2-y-z}$Co$_{z}$Se$_{2}$ in the samples investigated by EDX mapping were shown in Fig. 2 (a-c). The bright (colored) area is the area covered by each element. Se is uniformly distributed for all three samples while K, Fe, and Co display the pattern similar to the back-scattered electron image. This suggests that only K, Fe, and Co elements are responsible for phase separation. It is clear that the K and Fe concentrations are lower in the stripes (domains) than in the matrix while Co concentration is higher in the stripes than in the matrix. Hence, Co atoms prefer to enter into the stripe (domain) phase which is consistent with the report that Co substitution strongly suppresses superconductivity.\cite{TanD} The stripe(domain)-like brighter area maintains their shapes across the terraces created by the cleaving as shown by the marked ellipses in Fig. 2. This may suggest that the stripe(domain)-like brighter areas form three-dimensional spider-web-like network.\cite{DingX}

%TCIMACRO{\TeXButton{B}{\begin{table*}[tbp]\centering}}%
%BeginExpansion
\begin{table}[tbp]\centering%
%EndExpansion
\caption{Summary of measured compositions of K$_{x}$Fe$_{2-y-z}$Co$_{z}$Se$_{2}$ samples.}%
\begin{tabular}{ccccc}
\hline\hline
Nominal composition &&Measured&composition\\
K:Fe:Co:Se & K & Fe & Co & Se\\
\hline
1:1.8:0.2:2 & 0.79(5) & 1.37(3) & 0.06(0) & 2.\\
1:1.4:0.6:2 & 0.76(1) & 1.12(1) & 0.27(0) & 2\\
1:1:1:2 & 0.77(2) & 0.92(4) & 0.61(6) & 2\\
1:0.6:1.4:2 & 0.81(2) & 0.60(3) & 0.92(4) & 2\\
1:0.2:1.8:2 & 0.78(2) & 0.19(1) & 1.50(3) & 2\\
1:0:2:2 & 0.60(6) & 0 & 1.73(4) & 2\\
\hline\hline
\end{tabular}%
\label{EDX}%
%TCIMACRO{\TeXButton{E}{\end{table*}}}%
%BeginExpansion
\end{table}%
%EndExpansion

The average stoichiometry was measured by EDX for several single crystals in the same batch with multiple measuring points. The results indicate that the crystals are homogeneous within the scale of around 1$\times$1$\times$0.5 $mm^{3}$. The determined stoichiometries when fixing Se stoichiometry to be 2 are shown in Table I. Defects and vacancies of Fe and Co are observed for all investigated crystals which is common in A$_{x}$Fe$_{2-y}$Se$_{2}$ (A = K, Cs, Rb, Tl) compounds.\cite{GuoJ,LeiHC,WangDM} As the ratio of Co increases, the sum of Fe and Co ratio slightly increases while K ratio remains almost constant, similar to Ni doped K$_{x}$Fe$_{2-y}$Se$_{2}$ series.\cite{RyuH}

\begin{figure}
\centerline{\includegraphics[scale=0.42]{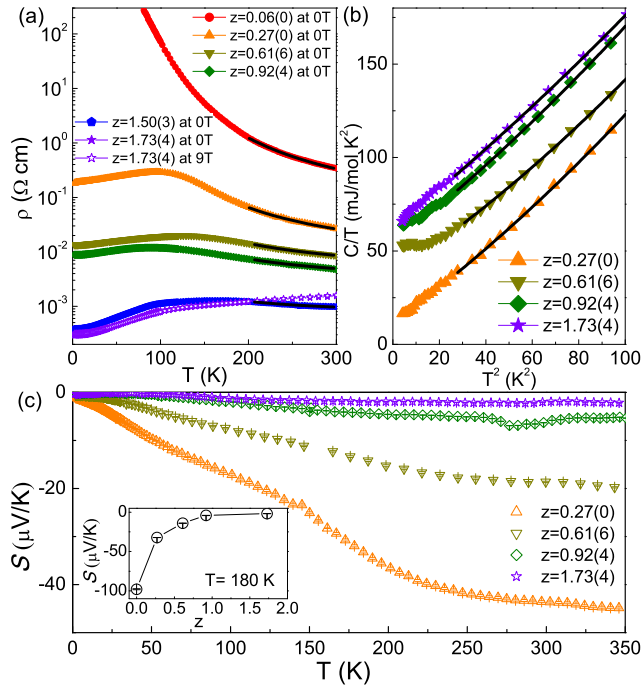}} \vspace*{-0.3cm}
\caption{(Color online) (a) Temperature dependence of the in-plane resistivity $\rho$(T) of K$_{x}$Fe$_{2-y-z}$Co$_{z}$Se$_{2}$ series at zero (closed symbols) and 9 T field (open symbols). The black solid lines are the fitted result using thermal activation model. (b) The relation between C/T and T$^{2}$ for K$_{x}$Fe$_{2-y-z}$Co$_{z}$Se$_{2}$ series at low temperature. The solid lines represent fits by the equation C/T=$\gamma$+$\beta_{3}$T$^{2}$+$\beta_{5}$T$^{4}$. (c) Temperature dependence of thermoelectric power S(T) for K$_{x}$Fe$_{2-y-z}$Co$_{z}$Se$_{2}$ series. The inset shows the thermoelectric power at T=180 K for different Co concentrations.}
\label{transport}
\end{figure}

As shown in Fig. 3 (a), 3.5\% of Co doping in K$_{x}$Fe$_{2-y}$Se$_{2}$ completely suppresses the superconductivity and results in an semiconducting $\rho$(T), consistent with previous study.\cite{TanD} As Co composition increases, there is a crossover from semiconductor to metallic state. Besides K$_{x}$Co$_{1.73(4)}$Se$_{2}$, all other crystals are metallic below a resistivity maximum $\rho_{max}$ and semiconducting above $\rho_{max}$ similar to K$_{x}$Fe$_{2-y}$Se$_{2-z}$S$_{z}$.\cite{LeiH107} High temperature part (above 200 K) of $\rho$(T) can be fitted by the thermal activation model $\rho=\rho_{0}exp(E_{a}/k_{B}T)$, where $\rho_{0}$ is a prefactor, $E_{a}$ is an activation energy, and $k_{B}$ is Boltzmann's constant [Fig. 3 (a)]. The obtained $\rho_{0}$ and $E_{a}$ are listed in Table II and are mostly smaller than for K$_{x}$Fe$_{2-y}$S$_{2}$ and KFe$_{0.85}$Ag$_{1.15}$Te$_{2}$.\cite{LeiH,LeiHC}

%TCIMACRO{\TeXButton{B}{\begin{table*}[tbp]\centering}}%
%BeginExpansion
\begin{table}[tbp]\centering%
%EndExpansion
\caption{Summary of $\rho_{0}$ values and activation energy, $E_{a}$, in K$_{x}$Fe$_{2-y-z}$Co$_{z}$Se$_{2}$.}%
\begin{tabular}{ccc}
\hline\hline
z & $\rho_{0}$($m\Omega$ $cm$) & $E_{a}$($meV$)\\
\hline
0.06(0) & 24.3(4) & 67.7(4)\\
0.27(0)& 4.20(9) & 47.1(5)\\
0.61(6)& 3.07(1) & 26.6(1)\\
0.92(4)& 2.15(3) & 21.4(3)\\
1.50(3)& 0.597(7) & 12.4(2)\\
\hline\hline
\end{tabular}%
\label{TAM}%
%TCIMACRO{\TeXButton{E}{\end{table*}}}%
%BeginExpansion
\end{table}%
%EndExpansion

Heat capacity of K$_{x}$Fe$_{2-y-z}$Co$_{z}$Se$_{2}$ series also exhibit the crossover from semiconductor to metal with Co increase, consistent with resistivity [Fig. 3(b)]. C/T-T$^{2}$ relations between 5 K and 10 K can be fitted by the formula C/T=$\gamma$+$\beta_{3}$T$^{2}$+$\beta_{5}$T$^{4}$. The Debye temperatures are obtained from $\Theta_{D}=(12\pi^{4}NR/5\beta)^{1/3}$, where N is the atomic number in the chemical formula and R is the gas constant. The obtained $\gamma$ values and Debye temperatures $\Theta_{D}$ for different Co ratios are listed in the Table III. Debye temperature for K$_{x}$Fe$_{2-y-z}$Co$_{z}$Se$_{2}$ series are similar suggesting that there are no considerable changes in atomic weight, structure, and bonding. Small $\gamma$ value for z=0.27(0) implies small density of states at the Fermi level similar to typical semiconductors while large $\gamma$ values for z$\geq$0.92(4) suggest accumulation of density of states as expected in metals.

%TCIMACRO{\TeXButton{B}{\begin{table*}[tbp]\centering}}%
%BeginExpansion
\begin{table}[tbp]\centering%
%EndExpansion
\caption{Summary of $\gamma$ values and Debye temperatures of K$_{x}$Fe$_{2-y-z}$Co$_{z}$Se$_{2}$}%
\begin{tabular}{ccc}
\hline\hline
z & $\gamma$(mJ mol$^{-1}$ K$^{-2}$) & $\Theta_{D}$(K)\\
\hline
0.27(0)& 11(1) & 220(4)\\
0.61(6)& 38(1) & 227(3)\\
0.92(4)& 54(1) & 215(3)\\
1.73(4)& 63(1) & 213(2)\\
\hline\hline
\end{tabular}%
\label{HC}%
%TCIMACRO{\TeXButton{E}{\end{table*}}}%
%BeginExpansion
\end{table}%
%EndExpansion

Thermoelectric power S(T) of K$_{x}$Fe$_{2-y-z}$Co$_{z}$Se$_{2}$ series shows negative values for all for different Co concentration which reveals that dominant carriers are electrons [Fig. 3(c)]. The magnitude of S(z) decreases as Co ratio increases to around 50\% [z=0.92(4)] and saturates [Inset in Fig. 3 (c)]. There are no obvious peaks in the thermoelectric power for K$_{x}$Fe$_{2-y-z}$Co$_{z}$Se$_{2}$ series between 2 K and 350 K suggesting that there are no dramatic Fermi surface changes.

\begin{figure}
\centerline{\includegraphics[scale=0.28]{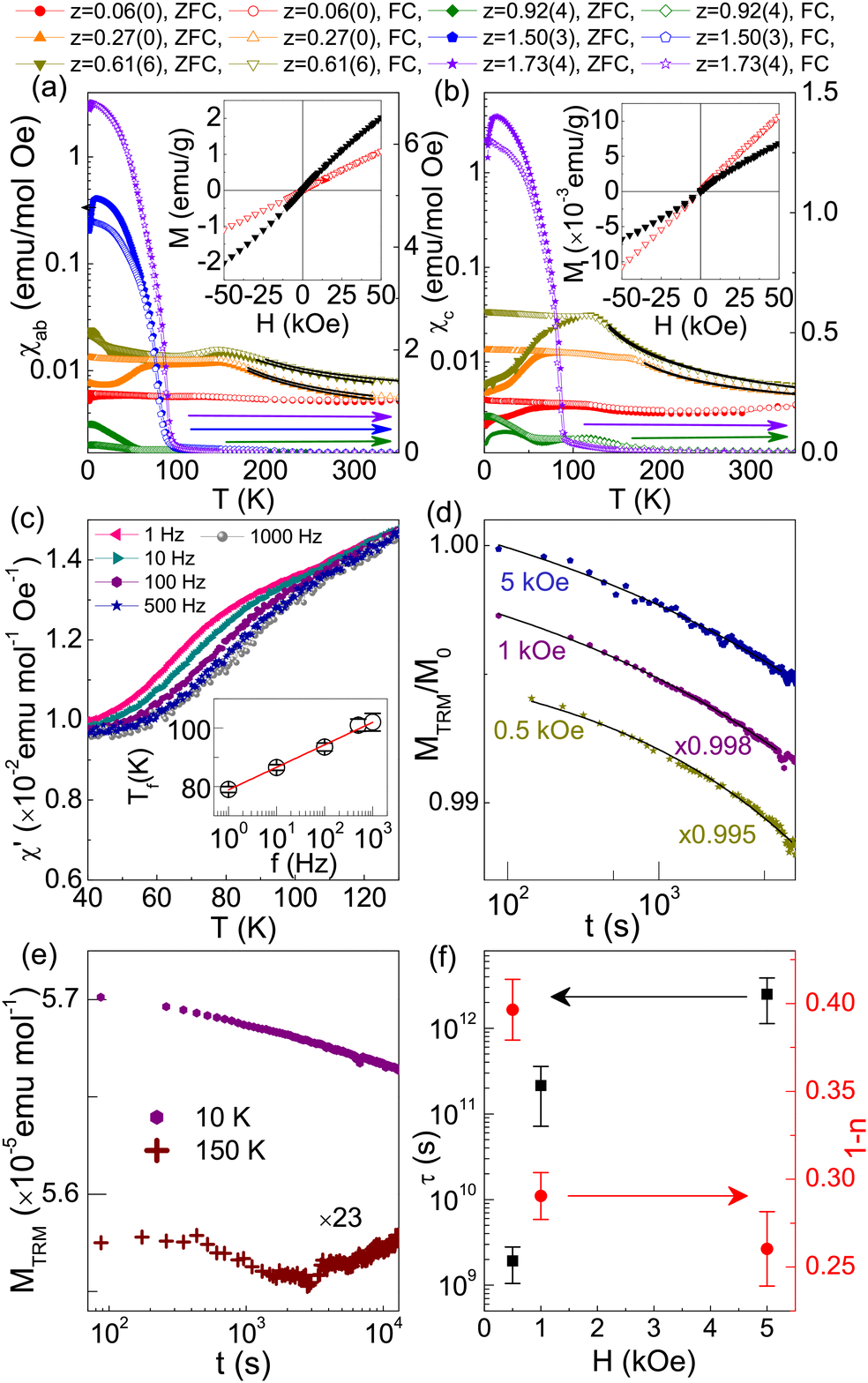}} \vspace*{-0.3cm}
\caption{(Color online) Temperature dependence dc magnetic susceptibilities of K$_{x}$Fe$_{2-y-z}$Co$_{z}$Se$_{2}$ series for (a) H$\bot$c and for (b) H$\|$c at H=1 kOe in ZFC and FC below 350 K. Inset figures of (a) and (b) are M-H curves of K$_{x}$Fe$_{2-y-z}$Co$_{z}$Se$_{2}$ with z=0.61(6) for H$\bot$c and H$\|$c, respectively at 1.8 K (black filled inverse triangle) and 300 K (red open inverse triangle). (c) Temperature dependence of ac susceptibility, $\chi'(T)$, measured at five different frequencies for z=0.27(0) of K$_{x}$Fe$_{2-y-z}$Co$_{z}$Se$_{2}$. Inset: Frequency dependence of $T_{f}$ with the linear fitting (solid line). (d) TRM vs. time for z=0.27(0) of K$_{x}$Fe$_{2-y-z}$Co$_{z}$Se$_{2}$ at 10 K and $t_{w}=100s$ with different dc field with fittings using stretched exponential function (solid lines). (e) $M_{TRM}$ vs. t at 10 K and 150 K with H = 1 kOe and $t_{w}=100s$. (f) H-field dependence $\tau(s)$ (black filled square) and 1-$n$ (red filled circle).}
\label{magnetism}
\end{figure}

Temperature dependent dc magnetic susceptibilities of K$_{x}$Fe$_{2-y-z}$Co$_{z}$Se$_{2}$ series show irreversible behaviors between zero-field-cooling (ZFC) and field-cooling (FC) at low temperature [Fig. 4 (a)]. This is a typical behavior of spin glass in magnetic field caused by the frozen magnetic spins in random directions below characteristic temperature T$_{f}$. Insets in Fig. 4(a) and Fig. 4(b) also suggest spin glass due to the linear field dependence of magnetic susceptibility with no hysteresis above T$_{f}$ (measured at 300 K) and s-shape loop of M-H curve below T$_{f}$ (measured at 1.8 K). Hence, 3.5\% of Co doping (z=0.06) not only suppresses superconductivity but also may result in spin glass. Ferromagnetic behavior appears when z$\geq$1.50(3).\cite{HuanG} High temperature regions (T $\geq$ 150 K) of K$_{x}$Fe$_{2-y-z}$Co$_{z}$Se$_{2}$ series follow Curie-Weiss law $\chi(T)=\chi_{0}+C/(T-\theta)$, where $\chi_{0}$ includes core diamagnetism, van Vleck and Pauli paramagnetism, $C$ is the Curie constant, and $\theta$ is the Curie-Weiss temperature [Fig. 4 (a) and (b)]. The obtained parameters are summarized in Table \ref{CW}. Negative $\theta$ values are observed even for crystals that order ferromagnetically, suggesting prevalent antiferromagnetic interactions probably come from the localized moment of the block (stripe) domain due to the phase separation.

%TCIMACRO{\TeXButton{B}{\begin{table*}[tbp]\centering}}%
%BeginExpansion
\begin{table}[tbp]\centering%
%EndExpansion
\caption{Summary of $\chi_{0}$, effective moment ($\mu_{eff}$), and $\theta$ values from Curie-Weiss fitting of K$_{x}$Fe$_{2-y-z}$Co$_{z}$Se$_{2}$}%
\begin{tabular}{cccccc}
\hline\hline
z & & $\chi_{0}$ & $\mu_{eff}$ & $\theta$ \\
  & &(emu mol$^{-1}$Oe$^{-1}$) & $(\mu_{B}$/Fe) & (K) \\
\hline
0.27(0) & H$\perp$C & 2.4(2)$\times$10$^{-3}$ & 2.33(8) &-91(5)\\
   & H$\parallel$C & 2.7(1)$\times$10$^{-3}$ & 1.88(6) & -118(3)\\
0.61(0) & H$\perp$C & 1.6(1)$\times$10$^{-3}$ & 2.76(3) & -98(1)\\
   & H$\parallel$C & 3.7(4)$\times$10$^{-3}$ & 2.79(9) & -81(9)\\
0.92(4) & H$\perp$C & 3.6(4)$\times$10$^{-3}$ & 2.7(2) & -120(9)\\
   & H$\parallel$C & 2.8(1)$\times$10$^{-3}$ & 2.38(2) & -146(1)\\
1.50(3) & H$\perp$C & 9.9(1)$\times$10$^{-3}$ & 1.42(4) & -218(2)\\
   & H$\parallel$C & 8.2(2)$\times$10$^{-3}$ & 0.65(9) & -257(6)\\
1.73(4) & H$\perp$C & 7.9(9)$\times$10$^{-4}$ & 3.2(4) & -86(9)\\
   & H$\parallel$C & 2.9(8)$\times$10$^{-3}$ & 1.8(6) & -138(9)\\

\hline\hline
\end{tabular}%
\label{CW}%
%TCIMACRO{\TeXButton{E}{\end{table*}}}%
%BeginExpansion
\end{table}%
%EndExpansion

Confirmation of spin glass comes from frequency dependence of the real part of ac susceptibility and thermoremanent magnetization. Frequency dependent susceptibility $\chi'(T)$ is shown in Fig. 4 (c). As frequency increases the characteristic temperature T$_{f}$ peak position increases whereas its magnitude decreases.\cite{MydoshJA} Frequency dependence of the peak shift is fitted by K=$\Delta$T$_{f}$/(T$_{f}$$\Delta$log$f$) (Fig. \ref{magnetism}), and the obtained K value is 0.036(1) in agreement with the canonical spin glass values (0.0045 $\leq$ K $\leq$ 0.08).\cite{MydoshJA} Thermoremanent magnetization (TRM) is shown in Fig. 4 (d). The sample was cooled down from 200 K (above T$_{f}$) to 10 K (below T$_{f}$) in four different magnetic fields, then kept at 10 K for $t_{w}=100s$. After that, the magnetic field was removed and M$_{TRM}$(t) was measured. As observed in Fig. 4(d), M$_{TRM}$(t) decays very slowly for all three different magnetic fields towards its non-zero equillibrium value.\cite{MydoshJA} On the other hand, M$_{TRM}$(t) measured at 150 K (above T$_{f}$) relaxes quickly in a short time (less than $\sim$100 s) [Fig. 4 (e)]. Slow relaxation behavior is fitted well by stretched exponential function, M$_{TRM}$(t) $\sim$ $M_{0}exp[-(t/\tau)^{1-n}]$, where $M_{0}$, $\tau$, and 1-$n$ are the glassy component, the relaxation characteristic time, and the critical exponent, respectively. As shown in Fig. 4 (f), the obtained $\tau$ is increases as H-field increases while $1-n$ stays close to 1/3 consistent with theoretical and experimental results for spin glass system.\cite{CampbellIA,ChuD}

\begin{figure}
\centerline{\includegraphics[scale=0.5]{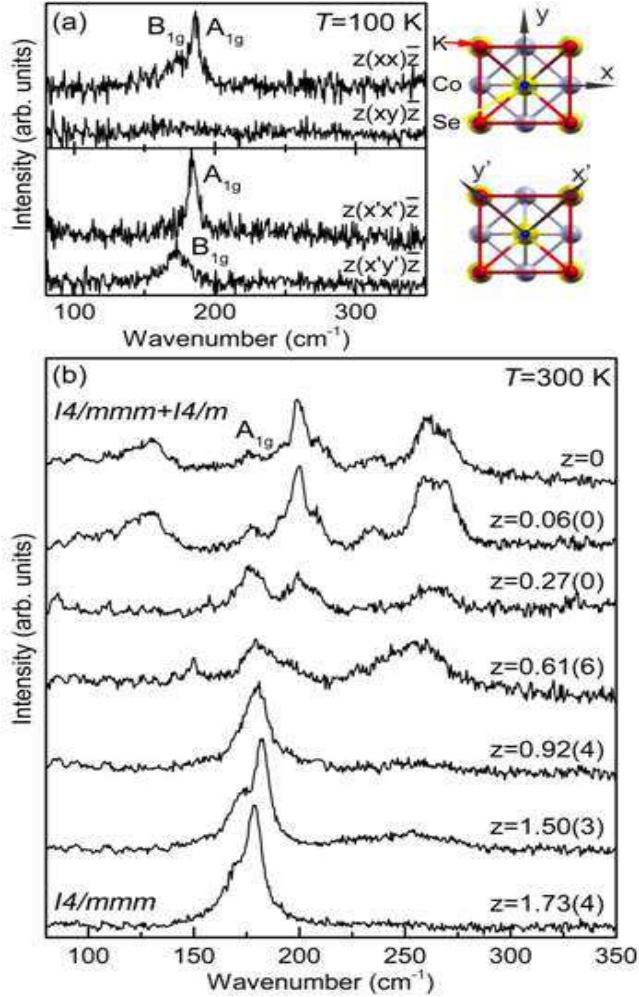}} \vspace*{-0.3cm}
\caption{(Color online) (a) Raman scattering spectra of K$_{0.6}$Co$_{1.73}$Se$_2$ single crystal in various scattering configurations ($\textbf{x}=[100], \textbf{y}=[010], \textbf{x'} = 1/\sqrt{2}[110], \textbf{y'}=1/\sqrt{2} [1\bar{1}0]$). (b) Raman scattering spectra of K$_x$Fe$_{2-y-z}$Co$_z$Se$_2$, [0 $\leq$ $z$ $\leq$ 1.73(4)] single crystals measured at room temperature from the (001)-plane of the samples.}
\label{Raman}
\end{figure}

Figure 5(a) shows polarized Raman scattering spectra of K$_{0.6}$Co$_{1.73}$Se$_2$ single crystal measured from the (001)-plane for the two sample orientations at 100 K using Jobin Yvon T64000 Raman system. According to selection rules for the $I4/mmm$ space group, peaks at about 174 and 184 cm$^{-1}$ (at 100 K) are assigned as B$_{1g}$ and A$_{1g}$ Raman modes, respectively.\cite{Lazarevic1,Lazarevic2}

Unpolarized Raman scattering spectra of K$_x$Fe$_{2-y-z}$Co$_z$Se$_2$ single crystals are presented in Fig. 5(b). For $z=1.73(4)$ samples, only two peaks, which were assigned as A$_{1g}$ ($\sim$180 cm$^{-1}$) and B$_{1g}$ ($\sim$169 cm$^{-1}$) modes, can be observed in the Raman spectrum. These modes are also observed for $z=1.50(3)$ and $z=0.92(4)$ samples. In fact, the A$_{1g}$ mode can be observed in Raman spectra for all concentrations of cobalt, suggesting that superconducting/metallic K$_{x}$Fe$_{2}$Se$_{2}$ phase is present in all investigated samples. Energy of this mode does not change significantly by varying concentrations of Co, as well as for different transition metal ions.\cite{Lazarevic1,Lazarevic2} For the intermediate concentration $[0.61(6)\leq z \leq 1.50(3)]$ a broad structure around 250 cm$^{-1}$ has been observed, which probably originates from the crystalline disorder in semiconducting K$_{2}$Fe$_{4}$Se$_{5}$ phase. In general, high disorder may cause relaxation of the selection rules, resulting in the appearance of a broad asymmetric structures. With further decreasing the Co concentration ($z \leq$ 0.27(0)), large number of Raman modes can be clearly observed in the spectra in addition to A$_{1g}$ mode of the superconducting/metallic K$_{x}$Fe$_{2}$Se$_{2}$ phase. These modes originate from the lattice vibrations within the ordered low symmetry semiconducting K$_{2}$Fe$_{4}$Se$_{5}$ phase.\cite{Lazarevic1,Lazarevic3}

\begin{figure}
\centerline{\includegraphics[scale=0.3]{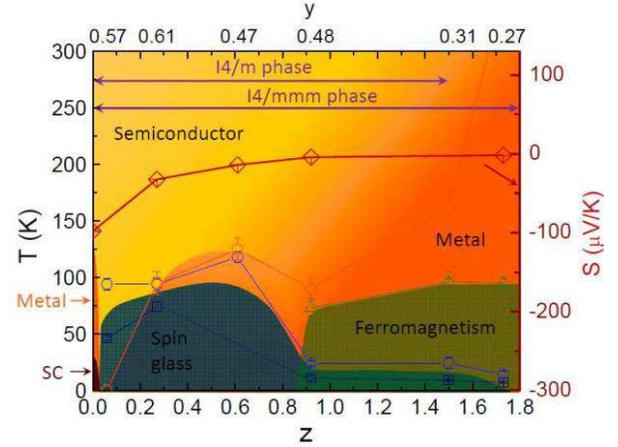}} \vspace*{-0.3cm}
\caption{(Color online) Magnetic and transport phase diagram. Open squares and open circles are spin glass characteristic temperature T$_{f}$ for H$\perp$c and H$\parallel$c direction, respectively. Open triangles are ferromagnetic transition temperature and open diamonds are thermoelectric power S$_{T}$ at 180 K. Open hexagons are resistivity maximums $\rho_{max}$ which show semiconductor [brighter (yellow) region] to metal [darker (orange) region] crossover. The lines at the top denote the regions of ordered I4/m and I4/mmm space groups.}
\label{PD}
\end{figure}

The magnetic and transport phase diagram of K$_{x}$Fe$_{2-y-z}$Co$_{z}$Se$_{2}$ series is presented in Fig. 6. When z$\sim$0 there is superconductivity below T$_{c}$$\sim$30 K, metallic resistivity below and semiconducting above about $\sim$125 K.\cite{GuoJ} By 3.5\% of Co-doping not only is superconductivity completely suppressed but also conductivity with emerging spin glass magnetic order below T$_{f}\sim$70 K in 1 kOe. A semiconducting/bad metal spin glass is found in proximity to superconducting state similar to copper oxides. As Co concentration increases, spin glass state is maintained while semiconductor to metal crossover is present at low temperatures up to z$\sim$0.6. After that, the spin glass and metallicity decrease up to z$\sim$0.9. With further increase in Co concentration, metallic conductivity spreads to higher temperature while spin glass is suppressed to lower temperature region (below $\sim$20 K) and ferromagnetic ground state emerges above the spin glass. K$_{x}$Co$_{z}$Se$_{2}$ with z=1.73(4) is a metal, consistent with previous reports.\cite{HuanG} We also note that ground state changes (Fig. 6) are concurrent with lattice parameter variations. Lattice parameters $a$ and $c$ for $I4/m$ show general drop as $z$ is increased, in contrast to lattice parameters of $I4/mmm$. When the ferromagnetism emerges, lattice parameter $c$ in $I4/mmm$ rapidly increases and is saturated similar to temperature dependence of Curie temperature.

It should be noted that both metallic conductivity and the total area of brighter stripe (block) regions increase with $z$. This could imply that the brighter stripe (block) area is metallic whereas the matrix is semiconducting, both with and without Co doping.\cite{TexierY,CharnukhaA} The z=0.06(0) crystal shows semiconducting behavior through entire temperature region we measured (1.8 K$\leq$ T $\leq$ 300 K), even though the metallic brighter stripe (block) areas are present (Fig. 2). This is most likely because the connectivity of the three dimensional metallic stripe (block) area is insufficient to create metallic percolation in the crystal.

The composite nature of our crystals and nano- to meso-scale mix of (super)conducting and semiconducting magnetic regions may also create states at interfaces.\cite{YanYJ,MukherjeeS} This somewhat complicates the physical interpretation of bulk measurements. However, since in K$_{x}$Fe$_{2-\delta}$Se$_{2}$ nanoscale phase separation exists below T$_{s}$ = 560 K,\cite{BaoW} most of conductivity changes below $T_{s}$ should come from the metallic regions. This is supported by recent angle-resolved photoemission results where orbital-selective Mott transition in K$_{x}$Fe$_{2-\delta}$Se$_{2}$ was observed above the crossover temperature.\cite{YiM} Therefore, the absolute values of resistivity and magnetization reflect the contribution of both semiconducting K$_{2}$Fe$_{4}$Se$_{5}$ and superconducting/metallic K$_{x}$Fe$_{2}$Se$_{2}$ parts of the crystal. Obtained thermoelectric power and heat capacity is contributed by metallic phase and semiconducting phase weighted by their conductivity. The estimated conductivity ratio between two phases are around $\sim$ 10$^{3}$ at 180 K which implies contribution of the metallic region is 1000 times larger than of semiconducting region. Co substitution in superconducting/metallic K$_{x}$Fe$_{2}$Se$_{2}$ unit cell is likely to have stronger effect on states associated with itinerant $d_{xz}/d_{yz}$ orbitals, for example via localization effect in an orbital-selective Mott localization scenario.\cite{YuRong,YuRong2} Further Co substitution and disorder might enhance conductivity by raising chemical potential and enlarging electron pockets, in agreement with our phase diagram.\cite{BerlijnT,CracoL,LuF}

\section{Conclusion}

We have demonstrated how the structure, phase separation, transport, and magnetic property evolve with Co doping concentration in K$_{x}$Fe$_{2-y-z}$Co$_{z}$Se$_{2}$ single crystals. A rich ground state phase diagram is discovered. By 3.5\% of Co doping superconductivity is suppressed while phase separation is still present which asserts the significance of arrangement and connectivity of phases for superconductivity. A semiconducting spin glass is discovered in close proximity to superconducting state in the phase diagram, similar to cooper oxides. Ferromagnetic metal state emerged above $\sim$50\% of Co concentration in agreement with the structural changes. The metallicity becomes dominant as the area of stipe (block) phases increases, however the connectivity of stripe phases may also be important for metallic conductivity.

\begin{acknowledgements}

M. Abeykoon and J. Hanson are gratefully acknowledged for experimental assistance at X7B beamline of NSLS at BNL. Work at Brookhaven is supported by the U.S. DOE under Contract No. DE-SC00112704 and in part by the Center for Emergent Superconductivity, an Energy Frontier Research Center funded by the U.S. DOE, Office for Basic Energy Science (K. W and C. P). This work was also supported by the Serbian Ministry of Education, Science and Technological Development under Projects ON171032 and III45018.

\end{acknowledgements}

\dag Present address: Advanced Light Source, E. O. Lawrence Berkeley National Laboratory, Berkeley, California 94720, USA

\S Present address: CNAM, Department of Physics, University of Maryland, College Park, Maryland 20742, USA

\end{document}